\documentclass{aastex61}

\shorttitle{SN II Light Curves and Spectra}
\shortauthors{Hicken et al.}

\begin{document}
\title{Type II Supernova Light Curves and Spectra From the C\MakeLowercase{f}A}

\author{Malcolm Hicken}
\affiliation{ Harvard-Smithsonian Center for Astrophysics, Cambridge, MA 02138, USA}
\author{Andrew S. Friedman}
\affiliation{ Harvard-Smithsonian Center for Astrophysics, Cambridge, MA 02138, USA}
\affiliation{Center for Theoretical Physics and Department of Physics, Massachusetts Institute of Technology, Cambridge, MA 02139}
\affiliation{ University of California, San Diego, La Jolla, CA 92093, USA}
\author{Peter Challis}
\affiliation{ Harvard-Smithsonian Center for Astrophysics, Cambridge, MA 02138, USA}
\author{Perry Berlind}
\affiliation{ Harvard-Smithsonian Center for Astrophysics, Cambridge, MA 02138, USA}
\author{Stephane Blondin}
\affiliation{ Aix Marseille Univ, CNRS, LAM, Laboratoire d'Astrophysique de Marseille, Marseille, France}
\author{Mike Calkins}
\affiliation{ Harvard-Smithsonian Center for Astrophysics, Cambridge, MA 02138, USA}
\author{Gil Esquerdo}
\affiliation{ Planetary Science Institute, 1700 East Fort Lowell Road, Tucson, AZ 85719, USA}
\author{Thomas Matheson}
\affiliation{ National Optical Astronomy Observatory, 950 North Cherry Avenue, Tucson, AZ, 85719, USA}
\author{Maryam Modjaz}
\affiliation{ Center for Cosmology and Particle Physics, New York University, 4 Washington Place, New York, NY 10003, USA}
\author{Armin Rest}
\affiliation{Space Telescope Science Institute, Baltimore, MD 21218, USA}
\author{Robert P. Kirshner}
\affiliation{ Harvard-Smithsonian Center for Astrophysics, Cambridge, MA 02138, USA}
\affiliation{ Gordon and Betty Moore Foundation, 1661 Page Mill Road, Palo Alto, CA 94304, USA}

\begin{abstract}

We present multiband photometry of 60 spectroscopically-confirmed supernovae (SN):  39 SN II/IIP, 19 IIn, one IIb and one that was originally classified as a IIn but later as a Ibn.  Forty-six have only optical photometry, six have only near infrared (NIR) photometry and eight have both optical and NIR.  The median redshift of the sample is 0.016.  We also present 192 optical spectra for 47 of the 60 SN.  All data are publicly available.  There are 26 optical and two NIR light curves of SN II/IIP with redshifts $z > 0.01$, some of which may give rise to useful distances for cosmological applications.  All photometry was obtained between 2000 and 2011 at the Fred Lawrence Whipple Observatory (FLWO), via the 1.2m and 1.3m PAIRITEL telescopes for the optical and NIR, respectively.  Each SN was observed in a subset of the $u'UBVRIr'i'JHK_s$ bands.  There are a total of 2932 optical and 816 NIR light curve points.  Optical spectra were obtained using the FLWO 1.5m Tillinghast telescope with the FAST spectrograph and the MMT Telescope with the Blue Channel Spectrograph.  Our photometry is in reasonable agreement with other samples from the literature.  Comparison with Pan-STARRS shows that two-thirds of our individual star sequences have weighted-mean $V$ offsets within $\pm$0.02 mag.  In comparing our standard-system SN light curves with common Carnegie Supernova Project objects using their color terms, we found that roughly three-quarters have average differences within $\pm$0.04 mag.  The data from this work and the literature will provide insight into SN II explosions, help with developing methods for photometric SN classification, and contribute to their use as cosmological distance indicators. 

\end{abstract}


\section{Introduction}

This paper presents CfA SN II light curves and spectra, all of which are publicly available from the journal, the CfA SN Group website\footnote{www.cfa.harvard.edu/supernova/index.html} and The Open Supernova Catalog.\footnote{https://sne.space/about/}  It is our hope that these data will be of use to the broader SN community for use in SN II analysis and cosmological calculations. 

Massive stars ($M \gtrsim 8M_\odot$) end their lives as core-collapse supernovae (CCSN).  Those that have retained a large portion of their hydrogen envelope are known as Type II SN, with spectra dominated by Balmer features.  Those that have lost their hydrogen envelope and have no Balmer lines but do have helium features are known as Type Ib.  Finally, those that have also lost much or all of their helium envelope and show no helium spectral features are known as Type Ic.  For reviews of these classifications, see, for example, \citet{galyam} and the introductions of \citet{smartt}, \citet{vandyk}, \citet{modjaz16} and \citet{liu16}.  A continuum seems to exist across the CCSN classes, depending on how much of the outer layer(s) has been lost.  SN II with thick hydrogen envelopes have a long plateau phase with roughly constant or slowly declining luminosity (SN IIP) whereas those with thinner envelopes decline more quickly (SN IIL) \citep[see, for example:][]{anderson, sanders}.  \citet{gall15} point out that, in addition to thinner envelopes, SN IIL need larger progenitor radii than SN IIP to give rise to their observed luminosities.  There also appears to be a continuous range of decline rates joining SN IIP and IIL, giving further weight to the continuum between CCSN types.  However, see \citet{valenti16} for more on the debate over SN IIP and IIL and \citet{morozova17} for discussion on whether there is a physical process that smoothly gives rise to a continuum between SN IIP and IIL or whether there is some specific mechanism that more abruptly gives rise to their differences.  Type IIb are an intermediate class between SN II and Ib.  They have hydrogen features in their early spectra but these quickly disappear as the SN ages, suggesting a thin hydrogen layer in the progenitor.  \citet{liu16} showed that there is a continuum of H$\alpha$ strengths between SN IIb and Ib.  Another kind of SN II is the Type IIn, distinguished by its narrow Balmer emission lines that are believed to arise from the SN blast wave colliding with previously-ejected progenitor material from stellar winds or eruptions.

SN II are of interest for several reasons.  They mark the death of many massive stars and play an important role in the chemical enrichment of the Universe.  SN II light curves and spectra give insight into SN II explosion mechanisms and progenitor properties.  SN II also serve as accurate distance indicators.   SN II are not as luminous as SN Ia \citep[see, for example,][]{richardson02} but do provide an alternative and independent means of measuring cosmological distances \citep[see, for example,][]{dejaeger}.  As more powerful telescopes, such as the Large Synoptic Survey Telescope,\footnote{www.lsst.org/} observe larger numbers of SN II at both low and high redshift, SN II will become more useful cosmological tools than in the past, offering an independent source of distances for calculating the expansion history of the universe.  Since most of the SN observed with large surveys will not be spectroscopically identified, it is important to understand the photometric nature of all types of SN so that sufficiently pure subsets (e.g., consisting only of SN Ia or SN IIP) can be photometrically separated and used for cosmology.
  
The potential cosmological use of SN II was pioneered by \citet{kirshner74, kirshner75} with the expanding photospheric method for measuring distances.  \citet{schmidt92} applied this method to 10 SN II to calculate their distances and a value for the Hubble constant.  \citet{schmidt94} improved their method and applied it to 18 SN II to find a Hubble constant of $H_o = 73 \pm 6(\mathrm{stat}) \pm 7(\mathrm{syst})$ km/s/Mpc, consistent within the error bars with more recent measurements such as the SN Ia and Cepheid-based value of $H_o = 73.24 \pm 1.74$ km/s/Mpc from \citet{riess16} and the $Planck$ cosmic microwave background radiation-based value of $H_o = 67.8 \pm 0.9$ km/s/Mpc from the \citet{planck16}.  

More recently, several optical and NIR SN II data sets have been published while other unpublished data sets have been used for analysis or cosmological applications.  We present a summary of many of them and their findings, in largely chronological order.  This may also assist anyone interested in compiling SN II data from the literature.  

\citet{poznanski09} combined optical light curves and spectra of 17 new SN IIP with those of 23 from the literature to find a Hubble diagram dispersion of 0.38 mag, which reduced to 0.22 mag when three $>3\sigma$ outliers were removed.  \citet{dandrea} presented light curves and spectra of 15 SDSS SN IIP and combined them with others from the literature to find a Hubble diagram dispersion of 0.29 mag for the combination.  

NIR photometry of SN II has also been produced and analyzed.  Dust extinction at NIR wavelengths is diminished relative to the optical, and there is the additional potential for smaller intrinsic dispersion in NIR light curves of SN II.  \citet{maguire10} explain that reduced extinction in the NIR gives rise to a lower error in the extinction estimate and has less of an effect on the fit between expansion velocity and NIR luminosity.  They also point out that SN IIP plateau-phase spectra have far fewer lines in the NIR than the optical.  They thus presume that estimates of SN IIP NIR luminosity should be less affected by variations in line strengths and line widths between different SN IIP.  They examined 12 SN IIP that had spectra and both $VI$ and NIR light curves.  However, only one of their SN was in the Hubble flow ($cz>3000$ km/s), suggesting the Hubble diagram dispersion they found would be larger than that of Hubble flow objects.  In the optical, they found a dispersion of 0.56 mag by combining $V$-band photometry and an estimate of the expansion velocity at +50d post-explosion.  In $I$, they found a dispersion of 0.5 mag.  \citet{maguire10} confirmed their hopes for the NIR by measuring a $J$ band dispersion of 0.39 mag, $\sim0.1$ mag lower than in $I$, suggesting that NIR light curves of Hubble-flow SN II would offer a similar improvement when compared to the optical.

In the past year, \citet{rodriguez16} presented preliminary results from a set of 16 SN II, showing a 0.12 mag dispersion Hubble diagram in each of the $JHK_s$ bands, better than the $BVI$ dispersions of 0.23, 0.17 and 0.19 mag, respectively, providing evidence that the smaller dispersion seen by \citet{maguire10} for very-nearby NIR SN II light curves would also be found in the Hubble flow.

More data was published by various groups in the following years.  Returning to the optical, \citet{arcavi} produced 21 SN II light curves from the Caltech Core Collapse Project.  \citet{taddia} published light curves and spectra of five CSP SN IIn.  \citet{faran} presented light curves and spectra of 23 SN IIP from LOSS and analyzed them.  \citet{anderson} provided V-band light curves and analysis for 116 SN II from the CSP and its predecessors (CT, C\&T, SOIRS and CATS).  They found evidence suggesting a continuum, where low-luminosity SN II have flat light curves during the plateau phase and higher-luminosity SN II have faster decline rates.  \citet{anderson} measured a dispersion of 0.56 mag around the  relation between the plateau-phase decline rate and peak magnitudes, suggesting that SN II can be used as pure photometric distance indicators.  \citet{galbany} published all of the bands for 51 of these 116 SN II while the CSP portion is not yet published.  \citet{gutierrez} presented an in-depth analysis of spectra of 123 SN II from the CSP and its predecessors, finding evidence suggesting that differences between SN IIP and IIL are related to the pre-explosion hydrogen envelope mass and do not come from different progenitor families.

\citet{rodriguez14}  found a very promising dispersion of 0.12 mag by using 13 SN II in the Hubble flow that have a well-constrained shock breakout epoch.   \citet{gonzalezgaitan} studied the rise times of 223 SN II light curves from SDSS and SNLS and found evidence that the early light curves of most SN II are powered by cooling of shock-heated ejecta.  They also found that massive hydrogen envelopes are indeed needed to explain the plateaus of SN II.  \citet{sanders} analyzed 76 multiband Pan-STARRS1 (PS1) SN II light curves and concluded that there does not appear to be two unconnected subclasses of SN II (IIP and IIL) but rather a one-parameter family likely related to the original mass of the progenitor, similar to the findings of \citet{anderson}.  \citet{sanders} further confirmed that SN IIP appear to  be standardizable with an intrinsic dispersion as low as $\sim$0.2 mag.  \citet{rubin} published 57 R-band SN II light curves from the Palomar Transient Factory.  \citet{valenti16} presented photometry of 12 SN II and combined them with well-sampled light curves from the literature to search for correlations between the slope of the linear light-curve decay and other properties.

In the past year, \citet{dejaeger} presented a Hubble diagram based on 73 SN II with a redshift range of $0.01 \leq z \leq 0.50$ from the CSP, SDSS and SNLS and found a Hubble diagram dispersion of 0.35 mag by applying the Photometric Colour Method with no redshift information, showing that SN II can be used as pure photometric distance indicators.  On a smaller sample of 61 SN II and using spectroscopic information, \citet{dejaeger} measured a dispersion of 0.27 mag.  Finally, \citet{gall17} presented light curves and spectra of nine SN IIP/L in the redshift range of $0.045 \leq z \leq 0.335$ and combined them with data from the literature to update previous EPM and SCM Hubble diagrams.  An interesting finding is that their three SN IIL are indistinguishable from their SN IIP in both the EPM and SCM Hubble diagrams.  Larger samples are needed to confirm this but it suggests that SN IIL may be useful cosmological distance indicators as well.  Given their higher luminosity than SN IIP, this could make SN IIL easier to find at higher redshifts, assuming a sufficient observing cadence is used to account for their faster decline.

\begin{deluxetable}{llllcccr}
\tablecolumns{8}
\tablewidth{0pc}
\tabletypesize{\tiny}
\tablecaption{SN II Discovery Data \label{table_snlist}}
\tablehead{ \colhead{SN} & \colhead{Type} & \colhead{Host Galaxy} &  \colhead{$z_{helio}$} & \colhead{Discovery Ref.} & \colhead{Opt LC} & \colhead{NIR LC} & \colhead{\# Spectra}}
\startdata
SN2000eo & IIn & MCG -2-9-3 & 0.010347 & IAUC 7524 & y & - & 14\\
SN2001ez & II & CGCG 329-009\tablenotemark{1} & 0.012916 & IAUC 7736 & y & - & 1\\
SN2001fa & IIn & NGC 673 & 0.017285 & IAUC 7737 & y & - &  5\\
SN2002bx & II & IC 2461 & 0.007539 & IAUC 7864 & y & - &  8\\
SN2002em & II & UGC 3430 & 0.013539 & IAUC 7955 & y & - & 0 \\
SN2005ay & II & NGC 3938 & 0.002699 & CBET 128 & - & y & 0\\
SN2005kd & IIn & CGCG 327-013\tablenotemark{1} & 0.015040 & IAUC 8630 & y & - & 0\\
SN2006at & II & Anon Gal & 0.015000 & CBET 424 & y & - & 1\\
SN2006be & II & IC 4582 & 0.007155 & CBET 449 & y & - &  3\\
SN2006bl & II & MCG +02-40-9 & 0.032382 & CBET 462 & y & - &  1\\
SN2006bo & IIn & UGC 11578 & 0.015347 & CBET 468 & y & - &  0\\
SN2006bv & IIn & UGC 7848 & 0.008382 & CBET 493 & y & - &  1\\
SN2006ca & II & UGC 11214 & 0.008903 & IAUC 8707 & y & - &  1\\
SN2006cd & IIP & IC 1179 & 0.037116 & CBET 508 & y & - &  1\\
SN2006gy & IIn & NGC 1260 & 0.019190 & CBET 644 &  y & - & 2\\
SN2006it & IIP & NGC 6956 & 0.015511 & CBET 660 & y & - &  0\\
SN2006iw & II & Anon Gal & 0.030700 & CBET 663 & y & - & 0\\
SN2006ov & II & NGC 4303 & 0.005224 & CBET 756  & y & - & 1\\
SN2007Q & II & NGC 5888 & 0.029123 & CBET 821  & y & - & 2\\
SN2007T & II & NGC 5828 & 0.013581 & CBET 833  & y & - & 0\\
SN2007aa & II & NGC 4030 & 0.004887 & CBET 848  & y & y & 4\\
SN2007ad & II & UGC 10845 & 0.027506 & CBET 854  & y & - & 1\\
SN2007ah & II & UGC 2931 & 0.019170 & CBET 869  & y & - & 1\\
SN2007ak & IIn & UGC 3293 & 0.015634 & CBET 875  & y & - & 0\\
SN2007av & II & NGC 3279 & 0.004650 & CBET 901  & y & y & 1\\
SN2007ay & II & UGC 4310 & 0.014527 & CBET 905  & y & - & 0\\
SN2007bb & IIn & UGC 3627 & 0.020858 & CBET 912  & y & - & 1\\
SN2007be & II & UGC 7800 & 0.012515 & CBET 917  & y & - & 1\\
SN2007bf & II & UGC 9121 & 0.017769 & CBET 919  & y & - & 1\\
SN2007cd & II & NGC 5174 & 0.022799 & CBET 950  & y & - & 0\\
SN2007ck & II & MCG +05-43-16 & 0.026962 & CBET 970  & y & - & 1\\
SN2007cm & IIn & NGC 4644 & 0.016501 & CBET 973  & y & - & 1\\
SN2007ct & II & NGC 6944 & 0.014734 & CBET 988  & y & - & 1\\
SN2007cu & II\tablenotemark{2} & UGC 10214 & 0.031358 & CBET 988  & y & - & 0\\
SN2007hv & II & UGC 2858 & 0.016858 & CBET 1056  & y & - & 1\\
SN2007od & II & UGC 12846 & 0.005784 & CBET 1116  & y & - & 13\\
SN2007pk & IIn & NGC 579 & 0.016655 & CBET 1129  & y & - & 7\\
SN2007rt & IIn & UGC 6109 & 0.022365 & CBET 1148  & y & y & 3\\
SN2008B & IIn & NGC 5829 & 0.018797 & CBET 1194  & y & - & 3\\
SN2008F & IIP & MCG -01-8-15 & 0.018366 & CBET 1207  & y & - & 1\\
SN2008aj & IIn & MCG +06-30-34 & 0.024963 & CBET 1259  & y & - & 1\\
SN2008bj & II & MCG +08-22-20 & 0.018965 & CBET 1314  & y & - & 16\\
SN2008bn & II & NGC 4226 & 0.024220 & CBET 1322  & y & - & 11\\
SN2008bu & II & ESO 586-G2 & 0.022115 & CBET 1341  & y & - & 0\\
SN2008bx & II & Anon Gal & 0.008399 & CBET 1348 & y & - & 3\\
SN2008gm & IIn & NGC 7530 & 0.011728 & CBET 1549  & y & - & 1\\
SN2008if & II & MCG -01-24-10 & 0.011475 & CBET 1619  & - & y & 1\\
SN2008in & II & NGC 4303 & 0.005224 & CBET 1636  & y & y & 13\\
SN2008ip & IIn & NGC 4846 & 0.015124 & CBET 1641  & y & y & 11\\
SN2009K & IIb & NGC 1620 & 0.011715 & CBET 1663  & - & y & 0\\
SN2009N & IIP & NGC 4487 & 0.003456 & CBET 1670  & y & - & 1\\
SN2009ay & II & NGC 6479 & 0.022182 & CBET 1728  & y & y & 6\\
SN2009dd & II & NGC 4088 & 0.002524 & CBET 1764  & y & - & 9\\
SN2009kn & IIn & MCG -03-21-6 & 0.015798 & CBET 1997  & y & - & 4\\
SN2009kr & II & NGC 1832 & 0.006468 & CBET 2006  & - & y & 2\\
SN2010aj & IIP & MCG -01-32-35 & 0.021185 & CBET 2201  & y & - & 7\\
SN2010al & Ibn\tablenotemark{3} & UGC 4286 & 0.017155 & CBET 2207  & y & y & 8\\
SN2010bq & IIn & UGC 10547 & 0.030988 & CBET 2241  & y & y & 5\\
SN2011an & IIn & UGC 4139  & 0.016308 & CBET 2668  & - & y & 2\\
SN2011ap & IIn & IC 1277  & 0.023630 & CBET 2670  & - & y & 9\\
\enddata
\tablenotetext{1}{Host galaxy from NED}
\tablenotetext{2}{Type from http://w.astro.berkeley.edu/bait/2007/sn2007cu.html}
\tablenotetext{3}{Type from \citet{pastorello}}
\tablecomments{The SN Type, Host Galaxy and Discovery Reference columns come from www.cbat.eps.harvard.edu/lists/Supernovae.html, except as noted.  The redshifts, $z_{helio}$, are of the host galaxy and come from NED, except for 2006at (CBET 441), 2006iw (\citet{dandrea}) and 2008bx (CBET 1359).}
\end{deluxetable}

The Harvard-Smithsonian Center for Astrophysics (CfA) Supernova Group has been a source of data for nearby SN since 1993.  The primary focus has been SN Ia but numerous CCSN have been observed as well.  In addition to various  individual SN papers, the CfA1-CfA4 data sets include a total of 345 SN Ia multiband optical light curves \citep{Riess99, Jha06, Hicken09, Hicken12}.  Optical spectroscopy of over 400 SN Ia \citep{Matheson08, Blondin12}  have been published.  In the near-infrared, the CfAIR2 set consists of 94 SN Ia light curves \citep{Friedman15} with earlier versions of a subset published by \citet{Woodvasey08}.  The CCSN photometry we acquired up until 2011 was processed along with the SN Ia photometry when the CfA3, CfA4 and CfAIR2 data sets were produced:  61 optical and 25 near-infrared stripped-envelope light curves were presented in \citet{Bianco14} and 60 SN II/IIn/IIb/Ibn light curves are presented in this work.   Two additional NIR light curves have been published elsewhere:  SN~2010jl \citep{fransson14} and SN~2011dh \citep{marion14}.   One SN IIb NIR light curve was produced after \citet{Bianco14} and it is presented here.  Additionally, optical light curves, spectra and analysis of SN IIP 2005cs and 2006bp were presented in \citet{dessart08}.  \citet{Modjaz14} presented and analyzed optical spectra of 73 stripped-envelope CCSN while CfA SN II spectra for 47 of the 60 SN from the current paper are presented here.  Finally, the CfA5 data set is currently being produced and will include optical SN light curves of all types taken after the CfA4 era (which included SN discovered up to mid-2010) plus any earlier ones that were missing calibration or host-galaxy subtraction images when the CfA4 light curves were created.  Spectra from the CfA5-era will also be published.

Section 2 describes the data and reduction and section 3 provides select comparisons of photometry with other data sets for overlapping objects.

\section{Data Acquisition and Reduction}

The CfA SN II sample consists of 60 objects:  39 SN II/IIP, 19 IIn, one IIb and one that was originally classified as a IIn but later as a Ibn \citep[2010al; see, for example,][]{pastorello}.  Forty-six objects have only optical photometry, six have only NIR photometry and eight have both.   The median redshift is 0.016.  Optical spectra for 47 of the 60 SN are presented in this work.  See Table \ref{table_snlist} for information on each SN's type, host galaxy, redshift, discovery reference, optical or NIR photometry, and number of CfA optical spectra.  Most of the SN were discovered as part of targeted searches.

There are 54 objects in this work with optical photometry:  36 II/IIP, 17 IIn and one Ibn.  Measurements were made in the $u'UBVr'i'$ bands and consist of a total 2932 light curve points.  These data were acquired on the 1.2m telescope at the FLWO\footnote{http://linmax.sao.arizona.edu/FLWO/48/48.html} at the same time as the CfA3 and CfA4 SN Ia data sets.  Twenty-six SN II light curves were generated as part of the light curve production process that created the CfA3 set and 28 as part of the process that created the CfA4 set.   Five of the CfA3-era SN II light curves were observed on the 4Shooter camera\footnote{http://linmax.sao.arizona.edu/FLWO/48/OLD/4shccd.html} (hereafter referred to as CfA3$_\mathrm{4SH}$), and 21 on the KeplerCam\footnote{http://linmax.sao.arizona.edu/FLWO/48/kepccd.html} (hereafter referred to as CfA3$_\mathrm{KEP}$).  The 28 CfA4-era SN II light curves (hereafter referred to as CfA4) were all observed on the KeplerCam.  Since all of the optical photometry reported here was produced as part of the CfA3 and CfA4 processing campaigns, see \citet{Hicken09} and \citet{Hicken12} for greater details on the instruments, observations, photometry pipeline, calibration and host-galaxy subtraction used to create the CfA SN II light curves.  It should be noted that the passbands shifted during part of the CfA4 observing campaign, likely due to deposits or condensation on the KeplerCam, resulting in two sets of color terms for the CfA4 calibration:  period 1 and period 2, one before the shift and one after.   Anyone using the CfA4-era natural system light curves should be careful to use the appropriate-period passbands, which can be found at our website.\footnote{www.cfa.harvard.edu/supernova/index.html}  Also, the 1.2m primary mirror deteriorated during the course of the CfA4 observing, causing a sensitivity loss of about 0.6 mag in $V$. 

There are 14 SN with NIR photometry in this work:  seven SN II, five IIn, one IIb and one Ibn.  We obtained our $JHK_s$-band photometry with the robotic 1.3-m Peters Automated InfraRed Imaging TELescope (PAIRITEL) at FLWO \citep{bloom06}.  PAIRITEL was a refurbished version of the Two Micron All Sky Survey (2MASS) North telescope using the transplanted 2MASS South camera \citep{skrutskie06}. It was utilized from 2005-2013 as a dedicated NIR imager for follow-up of transients, including SN of all types discovered by optical searches \citep{bloom06, Bianco14, Friedman15}.  Our NIR observing strategy was described elsewhere \citep{Woodvasey08, friedman12, Bianco14, Friedman15}. In particular, see \citet{Friedman15} for a detailed discussion of our image reduction and photometry pipelines including mosaicking, sky subtraction, and host galaxy subtraction. Whenever possible, we used an error-weighted mean of light curves derived from multiple host galaxy template images to remove flux contamination at the SN position. Since PAIRITEL was already on the 2MASS system, the SN brightness in each field was determined with differential photometry using reference field stars from the 2MASS point source catalog \citep{cutri03}.  There are a total of 816 NIR light curve points.

Eleven of our 14 NIR light curves had host galaxy subtraction as described in \cite{Friedman15}, while 3 objects were sufficiently isolated from the host galaxy nucleus (SN~2010bq, SN~2011an, SN~2011ap) and forced DoPHOT photometry \citep{schechter93} was used at the SN position without template image subtraction.  The NIR light curve of SN 2005ay is presented here while its optical light curves and spectra will be presented in a separate, future paper.  The optical light curve of SN 2009K was presented in \citet{Bianco14} but the NIR light curve did not have final calibration and host galaxy images and so it is presented here.

We acquired 192 optical spectra for 47 of the 60 SN in this work.  Spectra were obtained using the FLWO 1.5m Tillinghast telescope with the FAST spectrograph \citep{Fabricant98} and the MMT Telescope with the Blue Channel Spectrograph.  The FAST spectra were taken using a 3$''$ slit with an atmospheric corrector.  The observations at the MMT were taken with a 1$''$ slit and a 300 line/mm grating at the parallactic slit position.  All Spectra were reduced and flux calibrated by a combination of standard IRAF and custom IDL procedures \citep{Matheson05}.  The 2-d spectra underwent flat-fielding, cosmic-ray removal and extraction into 1-d spectra.  Pairs of spectroscopic standard stars were obtained to provide flux calibration (with no second order contamination) and assist in the removal of telluric features.  For more detail, see \citet{Matheson08} and \citet{Blondin12}.  The journal of observations of the spectra are in Table \ref{table_spectra_journal}.

\begin{deluxetable}{ccccccc}
\tablecolumns{7}
\tablewidth{0pc}
\tabletypesize{\footnotesize}
\tablecaption{Journal of Spectroscopic Observations \label{table_spectra_journal}}
\tablehead{ \colhead{SN} & \colhead{HJD} & \colhead{Tel./Instr.\tablenotemark{1}} & \colhead{Range (\AA)} & \colhead{Disp (\AA/pix)} & \colhead{Airmass} & \colhead{Exp. (s)} }
\startdata
SN2007aa & 2454154.989 & FAST & 3556-7457 & 1.46 & 1.48 &  1200\\
SN2007aa & 2454156.897 & FAST & 3477-7415 & 1.47 & 1.19 &  1500\\
SN2007aa & 2454169.866 & FAST & 3479-7419 & 1.47 & 1.20 &  1200\\
SN2007aa & 2454185.878 & FAST & 3476-7420 & 1.47 & 1.33 &  1500\\
SN2007ad & 2454155.028 & FAST & 3555-7456 & 1.46 & 1.11 &  1500\\
\enddata
\tablenotetext{1}{Telescope and instrument used for this spectrum:  FAST:  FLWO1.5m+FAST;  MMTblue: MMT+Blue Channel}
\tablecomments{Only a portion is shown here for guidance regarding its form and content.  This table is available in its entirety in a machine-readable form in the online journal.  It is also available at the CfA Web site:  www.cfa.harvard.edu/supernova/index.html}
\end{deluxetable}

\begin{deluxetable}{cccccccccccccc}
\tablecolumns{14}
\rotate
\tablewidth{0pc}
\tabletypesize{\scriptsize}
\tablecaption{Standard System Star Sequences \label{table_compstar}}
\tablehead{ \colhead{SN} & \colhead{Star} & \colhead{RA(J2000)} &  \colhead{DEC(J2000)} & \colhead{$V$} & \colhead{$N_V$} & \colhead{$U-B$} & \colhead{$N_U$} & \colhead{$B-V$} & \colhead{$N_B$} & \colhead{$V-r'$} & \colhead{$N_{r'}$} & \colhead{$V-i'$} & \colhead{$N_{i'}$} }
\startdata
SN2007aa & 01 & 12:00:38.638 & -01:06:55.25 & 16.213(0.014) & 2 & -- & 0 & 1.238(0.058) & 2 & 0.538(0.023) & 2 & 1.043(0.023) & 2 \\
SN2007aa & 02 & 12:00:36.905 & -01:05:53.67 & 17.245(0.016) & 2 & -- & 0 & 0.556(0.021) & 2 & 0.149(0.009) & 2 & 0.285(0.011) & 2 \\
SN2007aa & 03 & 12:00:35.217 & -01:03:51.48 & 17.324(0.026) & 2 & -- & 0 & 0.648(0.055) & 2 & 0.197(0.037) & 2 & 0.366(0.011) & 2 \\
SN2007aa & 04 & 12:00:25.058 & -01:02:49.29 & 17.802(0.030) & 2 & -- & 0 & 0.625(0.044) & 2 & 0.124(0.027) & 2 & 0.334(0.023) & 2 \\
SN2007aa & 05 & 12:00:24.144 & -01:09:55.72 & 17.359(0.014) & 2 & -- & 0 & 0.533(0.032) & 2 & 0.138(0.011) & 2 & 0.276(0.009) & 2 \\
SN2007aa & 06 & 12:00:20.107 & -01:06:22.99 & 17.451(0.029) & 2 & -- & 0 & 0.527(0.039) & 2 & 0.306(0.068) & 2 & 0.436(0.030) & 2 \\
SN2007aa & 07 & 12:00:19.767 & -01:05:03.02 & 15.299(0.012) & 2 & -- & 0 & 0.615(0.019) & 2 & 0.161(0.018) & 2 & 0.308(0.022) & 2 \\
SN2007aa & 08 & 12:00:16.229 & -01:02:14.43 & 17.954(0.033) & 2 & -- & 0 & 0.736(0.047) & 2 & 0.246(0.013) & 2 & 0.504(0.010) & 2 \\
SN2007aa & 09 & 12:00:14.168 & -01:04:24.72 & 17.185(0.016) & 2 & -- & 0 & 0.636(0.017) & 2 & 0.177(0.017) & 2 & 0.325(0.009) & 2 \\
SN2007aa & 10 & 12:00:08.517 & -01:07:18.48 & 15.374(0.015) & 2 & -- & 0 & 1.406(0.022) & 2 & 0.574(0.014) & 2 & 1.255(0.011) & 2 \\
SN2007aa & 11 & 12:00:07.955 & -01:04:42.70 & 16.902(0.020) & 2 & -- & 0 & 1.349(0.018) & 2 & 0.555(0.012) & 2 & 1.085(0.012) & 2 \\
SN2007aa & 12 & 12:00:07.616 & -01:04:57.01 & 16.756(0.012) & 2 & -- & 0 & 0.685(0.015) & 2 & 0.196(0.022) & 2 & 0.357(0.009) & 2 \\
\enddata
\tablecomments{Only a portion is shown here for guidance regarding its form and content.  This table is available in its entirety in a machine-readable form in the online journal.  It is also available at the CfA Web site:  www.cfa.harvard.edu/supernova/index.html}
\end{deluxetable}

\subsection{Star Sequences, Light Curves and Spectra}

In Table \ref{table_compstar} we present the standard system optical star sequences for the CfA SN II photometry and in Tables \ref{table_natlightcurve} and \ref{table_stdlightcurve} the natural system and standard system optical light curves, respectively.  Note should be made of the seven CfA3-era SN II optical light curves that were well-removed from their host galaxies and did not require host-galaxy subtraction.  They can be identified in Tables \ref{table_natlightcurve} and \ref{table_stdlightcurve} as those that have N$_{\mathrm{host}}$=0 host subtractions.  The rest of the CfA3-era SN II light curves have N$_{\mathrm{host}}$=1.  All of the CfA4-era light curves were host-subtracted and have N$_{\mathrm{host}}\geq$1.  We remind the reader that the CfA4 period-1 and period-2 natural system passbands are different and so special care should be taken to use the correct period-1 or period-2 passband for the natural system CfA4-era SN II light curves for each point.  In the last column of Table \ref{table_natlightcurve}, KEP1 means it was taken on the KeplerCam during CfA4 period 1 and KEP2 signifies period 2.  CfA4-era data before 2009 August 15 (MJD=55058) is period 1 and CfA4-era data after is period 2.  The CfA4 Ia light curves \citep{Hicken12} usually had multiple host subtractions for each light curve data point, giving rise to multiple values and uncertainties for such a data point.  The median of the multiple values, which arise from the multiple subtractions for each data point, was chosen to be the light curve value for that date.  The uncertainty was created from two components:  the median of the photometric pipeline uncertainties for each light curve point was added in quadrature to the standard deviation of the multiple host-subtraction light curve values for that point.  However, the CfA3-era light curves did not have multiple host galaxy subtractions and thus their uncertainty consists only of the photometric pipeline uncertainty.  To ensure that the CfA3-era and CfA4-era SN II light curve uncertainties are comparable, we chose to only use the median of the photometric pipeline uncertainties for the CfA4-era SN II light curves in this work (as opposed to adding it in quadrature to the standard deviation of the multiple subtraction light curve values for a given data point).    

\begin{figure}[ht!]
\plotone{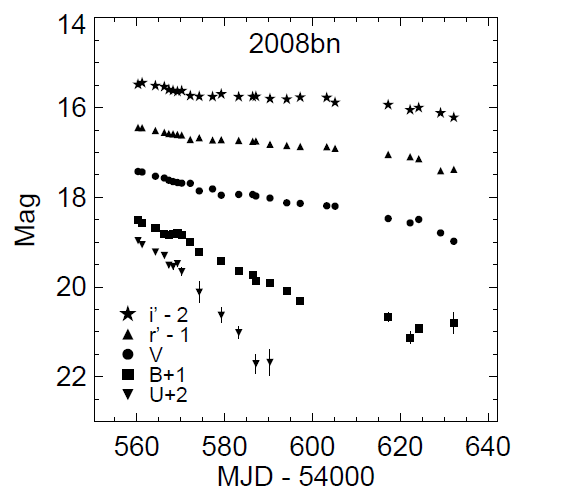}
\caption{Optical light curves of SN 2008bn.  Most error bars are smaller than the symbols.\label{snplot1}}
\end{figure}

\begin{figure}[ht!]
\plotone{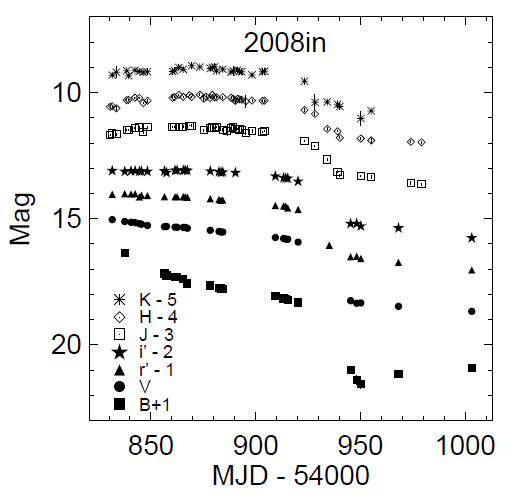}
\caption{Optical and NIR light curves of SN 2008in.  Most error bars are smaller than the symbols.\label{snplot2}}
\end{figure}

\begin{figure}[ht!]
\plottwo{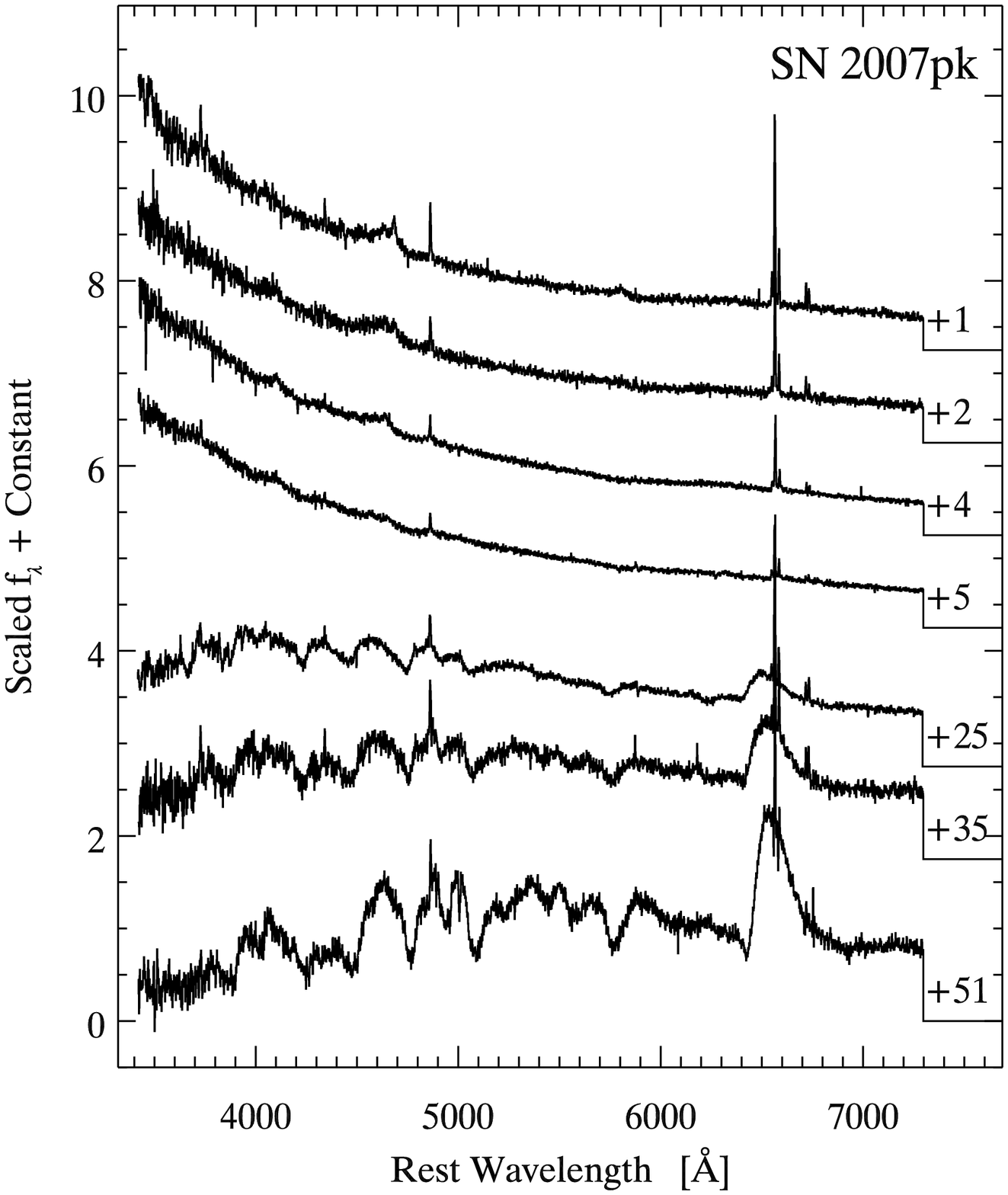}{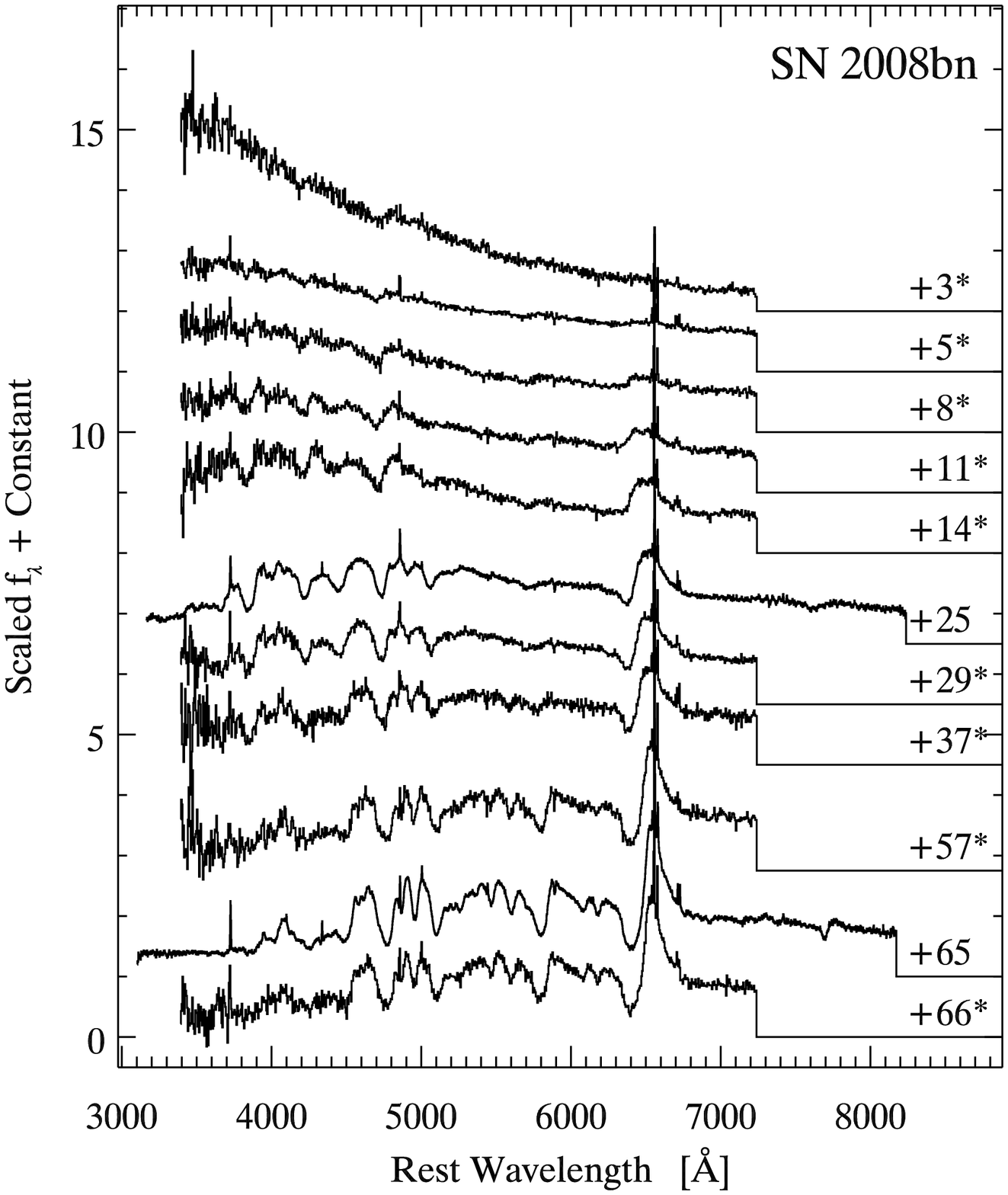}
\caption{Examples of SN~II spectral series from the CfA SN Group.  The flux units are $f_{\lambda}$ (erg s$^{-1}$ cm$^{-2}$ \AA$^{-1}$) that have been normalized and then additive offsets applied for clarity. The zero-flux level for each spectrum is marked with an extension on the red edge. The wavelength axis is corrected for the recession velocity of the host galaxy. The number associated with each spectrum indicates the age in (rest-frame) days from the date of discovery.
Spectra with low S/N have been binned to 5\,\AA\ per pixel; they are indicated with an asterisk appended to the age label. Shown are spectra of a Type IIn (SN~2007pk) and a Type IIP (SN~2008bn).\label{spectra}}
\end{figure}

In Table \ref{table_nirlightcurve} we present the PAIRITEL NIR light curves.  Figures \ref{snplot1} and \ref{snplot2} show two examples of CfA SN II light curves:  one optical (SN 2008bn) and one with both optical and NIR (SN 2008in).  Figure \ref{spectra} shows the spectral series of SN 2007pk, a IIn, and SN 2008bn, a IIP.  The photometry and spectra are available from the journal, the CfA SN Group website\footnote{www.cfa.harvard.edu/supernova/index.html} and The Open Supernova Catalog.\footnote{https://sne.space/about/}  The natural system passbands mentioned in \citet{Hicken12} are also available at our website.  The period-1 passbands can be  used for the CfA3 KeplerCam and the CfA4 period-1 natural system light curves while the period-2 passbands can be used for the CfA4 period-2 natural system light curves.  The 4Shooter natural system passbands can be found in \citet{Jha06} for use with the CfA3 4Shooter natural system light curves.

\begin{deluxetable}{cccccccc}
\tablecolumns{8}
\tablewidth{0pc}
\tabletypesize{\footnotesize}
\tablecaption{Natural-System SN Light Curves \label{table_natlightcurve}}
\tablehead{ \colhead{SN} & \colhead{Filter} & \colhead{MJD} & \colhead{N$_{\mathrm{host}}$} & \colhead{Mag} & \colhead{$\sigma$} & \colhead{Production Campaign} & \colhead{Camera}}
\startdata
SN2007aa & B & 54152.35648 & 0 & 16.167 & 0.016 & CfA3 & KEP \\
SN2007aa & B & 54153.32938 & 0 & 16.226 & 0.017 & CfA3 & KEP \\
SN2007aa & V & 54152.35348 & 0 & 15.703 & 0.012 & CfA3 & KEP \\
SN2007aa & V & 54153.32638 & 0 & 15.719 & 0.013 & CfA3 & KEP \\
SN2007aa & r' & 54152.35117 & 0 & 15.571 & 0.013 & CfA3 & KEP \\
SN2007aa & r' & 54153.32407 & 0 & 15.580 & 0.014 & CfA3 & KEP \\
SN2007aa & i' & 54152.34885 & 0 & 15.694 & 0.012 & CfA3 & KEP \\
SN2007aa & i' & 54153.32176 & 0 & 15.717 & 0.015 & CfA3 & KEP \\
SN2009N & B & 54857.50559 & 5 & 16.398 & 0.022 & CfA4 & KEP1\\
SN2009N & B & 54858.55471 & 5 & 16.401 & 0.024 & CfA4 & KEP1\\
SN2009N & V & 54857.51825 & 9 & 16.294 & 0.020 & CfA4 & KEP1\\
SN2009N & V & 54858.55843 & 5 & 16.303 & 0.023 & CfA4 & KEP1\\
SN2009N & r' & 54857.51585 & 4 & 16.272 & 0.028 & CfA4 & KEP1\\
SN2009N & r' & 54858.54485 & 5 & 16.261 & 0.016 & CfA4 & KEP1\\
SN2009N & i' & 54857.50628 & 8 & 16.192 & 0.027 & CfA4 & KEP1\\
SN2009N & i' & 54858.54786 & 5 & 16.159 & 0.015 & CfA4 & KEP1\\
\enddata
\tablecomments{This table presents the CfA natural-system SN II photometry.  N$_{\mathrm{host}}$ is the number of host-galaxy images subtracted from the same data image.  N$_{\mathrm{host}}$=0 means no host-galaxy subtraction was performed and the SN was sufficiently removed from the host galaxy.  $\sigma$ is the same as the light curve uncertainty as used in \citet{Hicken09} and is the same as $\sigma_{pipe}$ in \citet{Hicken12}.  The penultimate column lists during which light curve production campaign that data point was produced:  CfA3 or CfA4.  The last column lists which camera the SN data was acquired with.  For CfA4 data, a 1 or 2 is appended to 'KEP' to indicate if it was taken during period 1 or 2 and which set of natural-system bandpasses should be used.  Only the first two nights in each band of two SN (one from CfA3 and one from CfA4) are shown here for guidance regarding its form and content.  This table is available in its entirety in a machine-readable form in the online journal.  It is also available at the CfA Web site, as are the natural-system passbands:  www.cfa.harvard.edu/supernova/index.html}
\end{deluxetable}

\begin{deluxetable}{cccccc}
\tablecolumns{6}
\tablewidth{0pc}
\tabletypesize{\footnotesize}
\tablecaption{Standard-System SN Light Curves \label{table_stdlightcurve}}
\tablehead{ \colhead{SN} & \colhead{Filter} & \colhead{MJD} & \colhead{N$_{\mathrm{host}}$} & \colhead{Mag} & \colhead{$\sigma$} }
\startdata
SN2007aa & B & 54152.35648 & 0 & 16.197 & 0.016 \\
SN2007aa & B & 54153.32938 & 0 & 16.259 & 0.017 \\
SN2007aa & B & 54158.34955 & 0 & 16.378 & 0.021 \\
SN2007aa & B & 54158.36546 & 0 & 16.371 & 0.022 \\
SN2007aa & B & 54169.35950 & 0 & 16.552 & 0.018 \\
SN2007aa & V & 54152.35348 & 0 & 15.694 & 0.012 \\
SN2007aa & V & 54153.32638 & 0 & 15.709 & 0.013 \\
SN2007aa & V & 54158.34654 & 0 & 15.720 & 0.013 \\
SN2007aa & V & 54158.36245 & 0 & 15.729 & 0.012 \\
SN2007aa & V & 54159.46133 & 0 & 15.704 & 0.014 \\
SN2007aa & r' & 54152.35117 & 0 & 15.568 & 0.013 \\
SN2007aa & r' & 54153.32407 & 0 & 15.576 & 0.014 \\
SN2007aa & r' & 54158.34423 & 0 & 15.563 & 0.013 \\
SN2007aa & r' & 54158.36015 & 0 & 15.572 & 0.015 \\
SN2007aa & r' & 54159.45900 & 0 & 15.557 & 0.014 \\
SN2007aa & i' & 54152.34885 & 0 & 15.685 & 0.012 \\
SN2007aa & i' & 54153.32176 & 0 & 15.707 & 0.015 \\
SN2007aa & i' & 54158.34190 & 0 & 15.647 & 0.015 \\
SN2007aa & i' & 54158.35785 & 0 & 15.644 & 0.014 \\
SN2007aa & i' & 54159.45671 & 0 & 15.622 & 0.015 \\
\enddata
\tablecomments{This table presents the CfA standard-system SN II photometry.  N$_{\mathrm{host}}$ is the number of host-galaxy images subtracted from the same data image.  N$_{\mathrm{host}}$=0 means no host-galaxy subtraction was performed and the SN was sufficiently removed from the host galaxy.  $\sigma$ is the same as the light curve uncertainty as used in \citet{Hicken09} and is the same as $\sigma_{pipe}$ in \citet{Hicken12}.  Only the first five nights in each band of one SN are shown here for guidance regarding its form and content.  This table is available in its entirety in a machine-readable form in the online journal.  It is also available at the CfA Web site:  www.cfa.harvard.edu/supernova/index.html}
\end{deluxetable}

\begin{deluxetable}{ccccc}
\tablecolumns{5}
\tablewidth{0pc}
\tabletypesize{\footnotesize}
\tablecaption{NIR Natural-System SN Light Curves \label{table_nirlightcurve}}
\tablehead{ \colhead{SN} & \colhead{Filter} & \colhead{MJD} & \colhead{Mag} & \colhead{$\sigma$} }
\startdata
SN2005ay & J & 53480.15 & 14.513 & 0.011  \\
SN2005ay & J & 53481.21 & 14.513 & 0.027  \\
SN2005ay & J & 53482.17 & 14.461 & 0.022  \\
SN2005ay & J & 53486.18 & 14.455 & 0.013  \\
SN2005ay & J & 53488.17 & 14.529 & 0.044  \\
SN2005ay & H & 53480.15 & 14.319 & 0.034  \\
SN2005ay & H & 53481.21 & 14.474 & 0.034  \\
SN2005ay & H & 53482.17 & 14.437 & 0.027  \\
SN2005ay & H & 53486.18 & 14.278 & 0.018  \\
SN2005ay & H & 53487.17 & 14.252 & 0.022  \\
SN2005ay & K$_s$ & 53480.15 & 14.177 & 0.022  \\
SN2005ay & K$_s$ & 53481.21 & 14.208 & 0.035  \\
SN2005ay & K$_s$ & 53482.17 & 14.196 & 0.031  \\
SN2005ay & K$_s$ & 53486.18 & 14.102 & 0.015  \\
SN2005ay & K$_s$ & 53487.17 & 14.188 & 0.018  \\
\enddata
\tablecomments{This table presents the CfA NIR natural-system SN II photometry.  Only the first five nights in each band of one SN are shown here for guidance regarding its form and content.  This table is available in its entirety in a machine-readable form in the online journal.  It is also available at the CfA Web site:  www.cfa.harvard.edu/supernova/index.html}
\end{deluxetable}

\section{Photometry Comparison With Other Samples }

We compared the $BVr'i'$ CfA SN II star sequences and light curves with overlapping data from two other groups and find reasonable agreement overall, although a few objects have larger discrepancies.  

\vspace{5mm}

\subsection{Comparison of CfA and Pan-STARRS1 Star Sequences }

\citet{Scolnic15} (S15) introduced the Supercal method which provides a very useful analysis tool to estimate systematic photometric offsets among many of the SN Ia samples, including those from the CfA.  Their work enables the different samples to be placed much more reliably on one photometric system.  For anyone combining SN II samples from various groups and needing to put them on one system, we recommend that either the offsets from S15 be applied for photometric systems covered by S15 or that the general method be followed to calculate offsets for photometric systems not covered.

To check how accurate our star sequence calibrations are, we did not apply the Supercal method but simply compared our standard system star photometry with the recently released Pan-STARRS1 (PS1) Catalog\footnote{http://archive.stsci.edu/panstarrs/search.php?form=fuf}, converted to the standard system.  Using the coordinates for each of the stars from Table \ref{table_compstar}, we searched the PS1 catalog.  We required a separation of 0.5 arc seconds or less for a match.  We added the offsets from Table 3 of \citet{Scolnic15} to transform the PS1 catalog values (which have the calibration of S15) to the \citet{Tonry12} calibration.  We then applied the transformation equations of Table 6 of \citet{Tonry12} to convert to the standard system $BVri$.  Finally, we converted from SDSS\footnote{http://classic.sdss.org/dr7/algorithms/jeg\_photometric\_eq\_dr1.html} $ri$ to \citet{Smith} $r'i'$, though there is minimal difference.  Similar to S15, we removed PS1 stars brighter than 14.8, 14.9 and 15.1 mag in $gri$ to avoid saturated stars and used the broader of the two S15 color ranges ($0.3 < g-i < 1.0$) in order to have more stars to compare with.   After applying these conditions, 50 of the 54 CfA SN II optical star sequences still had some stars in common with the PS1 Catalog.  Additionally, we had a star sequence (but no light curve) for SN 2004et that we compared with PS1, bringing the number of star sequences compared with PS1 to 51.

The weighted means of the CfA-minus-PS1 matched-star differences are shown in Table \ref{table_cfa_ps1_sample} for the CfA SN II sample as a whole, and for the separate CfA3$_\mathrm{4SH}$, CfA3$_\mathrm{KEP}$ and CfA4 matched-star subsamples.  No effort was made to convert the CfA3$_\mathrm{4SH}$ $RI$ bands into $ri$ and so only $BV$ comparisons are available for CfA3$_\mathrm{4SH}$ stars.  The weighted-average differences of all the stars combined for the respective $BVr'i'$ bands are -0.018, 0.001, 0.012 and -0.014 mag.

\begin{deluxetable}{l|rrrr}
\tablecolumns{5}
\tablewidth{0pc}
\tabletypesize{\footnotesize}
\tablecaption{Comparison of CfA SN II Star Sequences with Pan-STARRS1 by CfA Subsample\label{table_cfa_ps1_sample}}
\tablehead{ \colhead{Sample} & \colhead{$\langle\Delta B \rangle_w$} & \colhead{$\langle\Delta V \rangle_w$} & \colhead{$\langle\Delta r' \rangle_w$} & \colhead{$\langle\Delta i'\rangle_w$}\\ 
\colhead{} & \colhead{(mmag)} & \colhead{(mmag)} & \colhead{(mmag)} & \colhead{(mmag)}}
\startdata
CfA SN II stars$ - $PS1 (all matches)            & -17.5 $\pm$1.4 & 1.0 $\pm$0.8 & 12.3 $\pm$1.0 & -14.1 $\pm$1.1 \\
\hline
CfA3$_\mathrm{4SH}$ SN II stars$ - $PS1 & -24.9 $\pm$6.9 &-0.4 $\pm$4.6 &                         &                           \\
S15 CfA3$_\mathrm{4SH} - $PS1                      & -34.5 $\pm$5.3 & 1.8 $\pm$3.8 &                         &                           \\
\hline
CfA3$_\mathrm{KEP}$ SN II stars$ - $PS1 & -11.8 $\pm$3.6 &12.3 $\pm$2.2 & 25.1 $\pm$2.8 &   0.0 $\pm$2.8  \\
S15 CfA3$_\mathrm{KEP} - $PS1                & -30.9 $\pm$5.3 &  2.6 $\pm$3.6 &  12.4 $\pm$3.5 &   0.8 $\pm$3.6  \\
\hline
CfA4$_\mathrm{KEP}$ SN II stars$ - $PS1 & -18.2 $\pm$1.6 &-0.8 $\pm$0.9 & 10.3 $\pm$ 1.1 & -16.5 $\pm$1.2  \\
S15 CfA4$_1 - $PS1  & -20.1 $\pm$5.7 &  4.5 $\pm$3.6 &  4.9 $\pm$3.3 &   -0.8 $\pm$3.6  \\
S15 CfA4$_2 - $PS1  &    5.2 $\pm$6.4 &  5.1 $\pm$3.7 &  9.0 $\pm$3.4 &   -1.4 $\pm$3.6  \\
\enddata
\tablecomments{This table presents the weighted mean of the difference between standard system CfA and Pan-STARRS1 matched stars (CfA minus PS1) in $BVr'i'$ for the whole CfA SN II sample and for the subsamples listed, subject to the luminosity and color restrictions mentioned in the text.  Below each subsample's line from this work is the S15 NGSL offset to serve as a rough comparison.  However, note that the quantities from this work and S15 are not directly comparable since they are different things, nor are they based on the same sample of stars, so any similarity or difference is more suggestive than conclusive.  Also, note that the signs for the S15 NGSL differences are reversed here since the order of subtraction is CfA minus PS1 here, versus PS1 minus CfA in S15.}
\end{deluxetable}

\begin{deluxetable}{lrrrrrrrr}
\tablecolumns{9}
\tablewidth{0pc}
\tabletypesize{\scriptsize}
\tablecaption{Comparison of CfA SN II Star Sequences with Pan-STARRS1 by SN\label{table_cfa_ps1_eachsn}}
\tablehead{ \colhead{SN} & \colhead{$\langle\Delta B \rangle_w$} & \colhead{N} & \colhead{$\langle\Delta V \rangle_w$} & \colhead{N} & \colhead{$\langle\Delta r' \rangle_w$} & \colhead{N} & \colhead{$\langle\Delta i'\rangle_w$} & \colhead{N}\\ 
\colhead{} & \colhead{(mag)} & \colhead{} & \colhead{(mag)} & \colhead{} & \colhead{(mag)} & \colhead{} & \colhead{(mag)}}
\startdata
SN2001ez & -0.059 $\pm$0.023 & 10 & -0.034 $\pm$0.016 & 10 &  &  &  & \\
SN2001fa & -0.028 $\pm$0.030 & 10 & 0.004 $\pm$0.021 & 10 &  &  &  & \\
SN2002bx & 0.014 $\pm$0.021 & 8 & 0.020 $\pm$0.014 & 8 &  &  &  & \\
SN2002em & -0.026 $\pm$0.008 & 11 & 0.000 $\pm$0.005 & 11 &  &  &  & \\
SN2004et & -0.007 $\pm$0.008 & 8 & 0.014 $\pm$0.005 & 8 & 0.016 $\pm$0.005 & 8 & -0.002 $\pm$0.005 & 8\\
SN2005kd & -0.074 $\pm$0.008 & 9 & -0.002 $\pm$0.005 & 9 & 0.023 $\pm$0.006 & 9 & -0.009 $\pm$0.006 & 9\\
SN2006at & -0.014 $\pm$0.030 & 4 & -0.011 $\pm$0.020 & 4 & 0.027 $\pm$0.027 & 4 & -0.002 $\pm$0.027 & 4\\
SN2006be & 0.026 $\pm$0.026 & 5 & 0.031 $\pm$0.017 & 5 & 0.065 $\pm$0.023 & 5 & 0.030 $\pm$0.024 & 5\\
SN2006bl & -0.015 $\pm$0.006 & 18 & 0.001 $\pm$0.004 & 18 & 0.015 $\pm$0.005 & 18 & -0.014 $\pm$0.005 & 18\\
SN2006bo & -0.001 $\pm$0.026 & 6 & -0.019 $\pm$0.018 & 6 & -0.001 $\pm$0.023 & 6 & -0.050 $\pm$0.024 & 6\\
SN2006bv & -0.023 $\pm$0.026 & 6 & -0.008 $\pm$0.018 & 6 & 0.020 $\pm$0.023 & 6 & -0.014 $\pm$0.024 & 6\\
SN2006ca & -0.013 $\pm$0.050 & 1 & -0.029 $\pm$0.035 & 1 & -0.017 $\pm$0.045 & 1 & -0.052 $\pm$0.047 & 1\\
SN2006cd & -0.004 $\pm$0.023 & 6 & -0.008 $\pm$0.016 & 6 & 0.026 $\pm$0.021 & 6 & -0.002 $\pm$0.021 & 6\\
SN2006gy & -0.028 $\pm$0.004 & 38 & -0.012 $\pm$0.002 & 38 & -0.008 $\pm$0.003 & 38 & -0.059 $\pm$0.003 & 38\\
SN2006it & -0.007 $\pm$0.005 & 27 & -0.019 $\pm$0.003 & 27 & -0.009 $\pm$0.004 & 27 & -0.050 $\pm$0.006 & 27\\
SN2006iw & -0.024 $\pm$0.048 & 2 & -0.003 $\pm$0.034 & 2 & 0.016 $\pm$0.043 & 2 & 0.003 $\pm$0.045 & 2\\
SN2006ov & -0.095 $\pm$0.035 & 1 & 0.023 $\pm$0.014 & 1 & 0.024 $\pm$0.017 & 1 & -0.021 $\pm$0.017 & 1\\
SN2007Q  & 0.066 $\pm$0.027 & 7 & 0.030 $\pm$0.018 & 7 & 0.043 $\pm$0.024 & 7 & -0.004 $\pm$0.025 & 7\\
SN2007T  & 0.117 $\pm$0.022 & 8 & 0.092 $\pm$0.015 & 8 & 0.088 $\pm$0.020 & 8 & 0.052 $\pm$0.020 & 8\\
SN2007aa & -0.011 $\pm$0.023 & 3 & 0.009 $\pm$0.015 & 3 & 0.023 $\pm$0.017 & 3 & -0.005 $\pm$0.015 & 3\\
SN2007ad & -0.035 $\pm$0.047 & 3 & -0.025 $\pm$0.033 & 3 & -0.034 $\pm$0.042 & 3 & -0.030 $\pm$0.044 & 3\\
SN2007ah & -0.014 $\pm$0.008 & 11 & -0.013 $\pm$0.005 & 11 & 0.001 $\pm$0.005 & 11 & -0.038 $\pm$0.005 & 11\\
SN2007av & -0.012 $\pm$0.009 & 7 & 0.021 $\pm$0.006 & 7 & 0.018 $\pm$0.011 & 7 & -0.008 $\pm$0.011 & 7\\
SN2007bb & 0.138 $\pm$0.026 & 5 & 0.111 $\pm$0.018 & 5 & 0.104 $\pm$0.023 & 5 & 0.048 $\pm$0.024 & 5\\
SN2007be & -0.013 $\pm$0.023 & 8 & 0.029 $\pm$0.016 & 8 & 0.046 $\pm$0.021 & 8 & 0.028 $\pm$0.021 & 8\\
SN2007bf & -0.022 $\pm$0.027 & 9 & 0.015 $\pm$0.018 & 9 & 0.043 $\pm$0.024 & 9 & 0.038 $\pm$0.025 & 9\\
SN2007cd & 0.003 $\pm$0.023 & 8 & 0.035 $\pm$0.016 & 8 & -0.057 $\pm$0.021 & 8 & -0.073 $\pm$0.021 & 8\\
SN2007ck & 0.008 $\pm$0.010 & 12 & 0.004 $\pm$0.005 & 12 & 0.032 $\pm$0.006 & 12 & 0.010 $\pm$0.007 & 12\\
SN2007cm & 0.041 $\pm$0.027 & 1 & 0.053 $\pm$0.016 & 1 & 0.078 $\pm$0.017 & 1 & 0.059 $\pm$0.017 & 1\\
SN2007ct & 0.032 $\pm$0.011 & 37 & 0.038 $\pm$0.006 & 37 & 0.048 $\pm$0.008 & 37 & 0.015 $\pm$0.007 & 37\\
SN2007cu & 0.022 $\pm$0.014 & 7 & 0.036 $\pm$0.007 & 7 & 0.054 $\pm$0.009 & 7 & 0.032 $\pm$0.009 & 7\\
SN2007hv & -0.053 $\pm$0.006 & 23 & -0.003 $\pm$0.003 & 23 & 0.010 $\pm$0.004 & 23 & -0.042 $\pm$0.004 & 23\\
SN2007od & -0.000 $\pm$0.031 & 3 & -0.007 $\pm$0.022 & 3 & 0.030 $\pm$0.028 & 3 & 0.010 $\pm$0.028 & 3\\
SN2007pk & -0.034 $\pm$0.010 & 20 & -0.013 $\pm$0.005 & 20 & -0.001 $\pm$0.008 & 20 & -0.029 $\pm$0.008 & 20\\
SN2007rt & -0.001 $\pm$0.011 & 6 & 0.014 $\pm$0.006 & 6 & 0.032 $\pm$0.008 & 6 & 0.011 $\pm$0.009 & 6\\
SN2008aj & -0.014 $\pm$0.014 & 4 & -0.008 $\pm$0.007 & 4 & 0.010 $\pm$0.009 & 4 & -0.011 $\pm$0.009 & 4\\
SN2008B  & -0.008 $\pm$0.026 & 4 & 0.015 $\pm$0.014 & 4 & 0.055 $\pm$0.017 & 4 & 0.041 $\pm$0.018 & 4\\
SN2008F  & 0.036 $\pm$0.013 & 7 & 0.038 $\pm$0.007 & 7 & 0.051 $\pm$0.009 & 7 & 0.024 $\pm$0.009 & 7\\
SN2008bj & 0.006 $\pm$0.010 & 6 & 0.028 $\pm$0.007 & 6 & 0.054 $\pm$0.008 & 6 & 0.018 $\pm$0.008 & 6\\
SN2008bn & -0.026 $\pm$0.010 & 7 & 0.001 $\pm$0.006 & 7 & 0.021 $\pm$0.007 & 7 & 0.004 $\pm$0.007 & 7\\
SN2008bu & -0.077 $\pm$0.006 & 17 & -0.023 $\pm$0.004 & 17 & -0.040 $\pm$0.005 & 17 & -0.042 $\pm$0.005 & 17\\
SN2008bx & -0.031 $\pm$0.011 & 11 & 0.016 $\pm$0.006 & 11 & 0.033 $\pm$0.007 & 11 & 0.003 $\pm$0.007 & 11\\
SN2008gm & -0.023 $\pm$0.010 & 6 & -0.008 $\pm$0.006 & 6 & 0.009 $\pm$0.007 & 6 & -0.027 $\pm$0.007 & 6\\
SN2008in & -0.057 $\pm$0.013 & 4 & -0.018 $\pm$0.008 & 4 & 0.023 $\pm$0.010 & 4 & -0.019 $\pm$0.009 & 4\\
SN2008ip & -0.015 $\pm$0.012 & 5 & 0.011 $\pm$0.007 & 5 & 0.034 $\pm$0.008 & 5 & 0.019 $\pm$0.008 & 5\\
SN2009ay & -0.013 $\pm$0.007 & 14 & -0.002 $\pm$0.004 & 14 & 0.011 $\pm$0.005 & 14 & -0.006 $\pm$0.005 & 14\\
SN2009dd & -0.024 $\pm$0.032 & 1 & 0.014 $\pm$0.027 & 1 & 0.048 $\pm$0.020 & 1 & 0.020 $\pm$0.017 & 1\\
SN2009kn & -0.001 $\pm$0.005 & 42 & 0.005 $\pm$0.003 & 42 & 0.011 $\pm$0.004 & 42 & -0.005 $\pm$0.004 & 42\\
SN2009N  & -0.019 $\pm$0.013 & 4 & 0.001 $\pm$0.008 & 4 & 0.011 $\pm$0.010 & 4 & -0.010 $\pm$0.010 & 4\\
SN2010aj & -0.016 $\pm$0.014 & 4 & 0.003 $\pm$0.008 & 4 & 0.029 $\pm$0.010 & 4 & 0.019 $\pm$0.010 & 4\\
SN2010al & -0.009 $\pm$0.006 & 19 & 0.006 $\pm$0.004 & 19 & 0.025 $\pm$0.004 & 19 & 0.015 $\pm$0.004 & 19\\
\enddata
\tablecomments{This table presents the weighted mean of the difference between standard system CfA and PS1 stars (CfA minus PS1) in BVr'i' for each individual SN's star sequence and how many stars were matched.  50 CfA SN II have star matches with PS1 after luminosity and color restrictions.  Four SN star sequences have no PS1 matches.  Additionally, we calibrated a star sequence for SN 2004et on the KeplerCam and include it in this analysis even though the light curve is not available.}
\end{deluxetable}

\begin{deluxetable}{lrrrrrrrr}
\tablecolumns{9}
\tablewidth{0pc}
\tabletypesize{\footnotesize}
\tablecaption{Comparison of CfA SN II Star Sequences with CSP by SN\label{table_cfa_csp_stars}}
\tablehead{ \colhead{SN} & \colhead{$\langle\Delta B \rangle$} & \colhead{N} & \colhead{$\langle\Delta V \rangle$} & \colhead{N} & \colhead{$\langle\Delta r' \rangle$} & \colhead{N} & \colhead{$\langle\Delta i'\rangle$} & \colhead{N}\\ 
\colhead{} & \colhead{(mag)} & \colhead{} & \colhead{(mag)} & \colhead{} & \colhead{(mag)} & \colhead{} & \colhead{(mag)}}
\startdata
SN2006be &  0.004 & 5   &  0.036 & 5   &  0.058 & 4   &  0.036 & 5\\
SN2006bl & -0.016 & 13   &  0.002 & 13   &  0.015 & 13   &  0.002 & 13\\
SN2006bo & -0.023 & 4   & -0.017 & 4   & -0.083 & 4   & -0.017 & 4\\
SN2006it & -0.025 & 12   & -0.015 & 12   & -0.019 & 12   & -0.015 & 12\\
SN2007aa &  0.016 & 10   &  0.017 & 10   &  0.043 & 10   &  0.017 & 10\\
SN2007od & -0.018 & 6   & -0.015 & 6   & -0.014 & 6   & -0.015 & 6\\
SN2008bu & -0.006 & 9   &  0.019 & 9   &  0.024 & 9   &  0.019 & 9\\
SN2009N &  0.029 & 8   &  0.039 & 8   &  0.051 & 7   &  0.039 & 8\\
\enddata
\tablecomments{This table presents the average of the difference between standard system CfA and CSP stars (CfA minus CSP) in BVr'i' for 8 individual SN star sequences and how many stars were matched.}
\end{deluxetable}

\begin{deluxetable}{lcrr}
\tablecolumns{4}
\tablewidth{0pc}
\tabletypesize{\footnotesize}
\tablecaption{Comparison of CfA SN II Light Curves with CSP by SN\label{table_cfa_csp_lc}}
\tablehead{ \colhead{SN} & \colhead{Band} & \colhead{$\langle\Delta mag \rangle$} & \colhead{N} }
\startdata
SN2006bo & B & -0.027 & 3\\
SN2006bo & V & -0.046 & 4\\
SN2006bo & r' & -0.033 & 4\\
SN2006bo & i' & -0.011 & 4\\
\hline
SN2006be & V & 0.071 & 7\\
SN2006bl & V & 0.106 & 5\\
SN2006it & V & -0.038 & 1\\
SN2007aa & V & -0.010 & 5\\
SN2007od & V & -0.015 & 6\\
SN2008bu & V & -0.033 & 1\\
SN2009N & V & 0.032 & 1\\
\enddata
\tablecomments{This table presents the average of the difference between standard system CfA and CSP light curve points for eight SN (CfA minus CSP, with $\Delta$MJD $<$ 0.6d), in BVr'i' for 2006bo and V for the other seven, and how many light curve points were matched.}
\end{deluxetable}

In this work, the CfA $B$ calibration is about 0.02 brighter, in the same direction found by S15 in their Table 4 (NGSL column).  It should be noted that the comparisons of this work and S15 are not the same.  First of all, the star sequences in S15 are different from those in this work.  More importantly, S15 used observed natural-system and synthetic natural-system photometry while this work is using standard system photometry so any similarities or differences in the values between the two works' offsets are not conclusive but merely suggestive.  

In this work, CfA $V$ has virtually no offset with the PS1 conversion to $V$ for the CfA3$_\mathrm{4SH}$ and CfA4 subsamples, in rough agreement with S15.  However, the CfA3$_\mathrm{KEP}$ subsample is 0.012 mag fainter, about 0.01 mag fainter than the offset S15 found.  

CfA $r'$ is fainter than PS1 in this work, as is the case in S15, though the offset here is greater than in S15.  There is excellent agreement between the CfA3$_\mathrm{KEP}$ $i'$ offsets in the two works but the CfA4 $i'$ offset in this work is 0.0165 mag brighter while there is virtually no offset in S15.  

With the exception of CfA3$_\mathrm{KEP}$ $V$ and CfA4 $i'$, there is reasonable directional agreement in the CfA-minus-PS1 offsets of this work and S15,  in the sense of both agreeing on which calibration is fainter or brighter.  

On a SN-by-SN basis, Table \ref{table_cfa_ps1_eachsn}  shows the weighted mean of the CfA-minus-PS1 star differences for each CfA SN II that had matches.  19 of the 51 SN II (37$\%$) have $V$ offsets within $\pm$0.01 mag, 34 (67\%) have $V$ offsets within $\pm$0.02 mag, 42 (82\%) within $\pm$0.03 mag, and 48 (94\%) within $\pm$0.038 mag.  These can be considered to have reasonable agreement with PS1.  However, there are 3 SN with larger calibration discrepancies.  SN 2007bb has a $V$ offset of 0.111 mag based on 5 matched stars, SN 2007T has a 0.092 mag offset based on 8 stars and SN 2007cm has a 0.053 mag offset based on only 1 star.  

\subsection{Comparison of CfA and CSP Star Sequences and SN II Light Curves}

The CSP provided us with standard-system $BVr'i$ star sequences for their upcoming SN II light curve publication (Carlos Contreras --- private communication).  Table \ref{table_cfa_csp_stars} shows the $BVr'i'$ comparisons between 8 CfA and CSP star sequences, with the average of the differences in $V$ ranging from -0.014 to 0.035 mag, showing reasonable agreement.  

We also took the CSP natural system $V$ light curves for 7 objects from \citet{anderson} and the natural system $BVr'i'$ light curves for SN 2006bo from \citet{taddia} and applied color terms\footnote{$CT_B = +0.069, CT_V = -0.063, CT_r = -0.016, CT_i= 0.0$ \citep[][and Carlos Contreras --- private communication]{stritzinger}} to convert them to the standard system and compare with the CfA standard system light curves from Table \ref{table_stdlightcurve}.  Applying star-derived color terms to SN photometry to create standard system light curves is fraught with danger and inaccuracy (the user is encouraged to use the natural system light curves whenever possible) so this is intended only as a rough comparison and sanity check.  Only points that are within 0.6d of each other are matched.  No interpolation was used to provide for more points of comparison.  Table \ref{table_cfa_csp_lc} shows the average of the light curve differences for the 8 SN II in common (11 comparisons in total:  8 in $V$ and 1 in each of $Br'i'$).  Of the 11 comparisons, eight are within $\pm$0.04 mag while one has CfA 0.046 mag brighter, another has CfA 0.071 mag fainter and the most disparate one has CfA 0.106 mag fainter.  This is comparable to the similarity that the CfA3 and CfA4 SN Ia light curves had with other groups' light curves in \citet{Hicken09} and \citet{Hicken12} and suggests that the CfA SN II light curves are of reasonable accuracy.

\section{Conclusion }

The CfA SN II light curve sample consists of 60 multiband  light curves:  46 have only optical, six have only NIR and eight have both optical and NIR.  We also present 192 optical spectra for 47 of the 60 SN.  This work includes a total of 2932 optical and 816 NIR light curve points.  There are 26 optical and two NIR light curves of SN II/IIP with redshifts $z > 0.01$, some of which may give rise to useful distances for Hubble diagrams.  Select comparisons with other groups' data show reasonable agreement in the vast majority of cases.  These light curves add to a growing body of SN II data from the literature.  This collective body of data (from the literature, this work and future samples) will be useful for providing greater insight into SN II explosions, developing analytic methods for photometric SN classification, and opening the path for their increasing use as cosmological distance indicators.  Having another method for measuring the expansion history of the Universe, independent from SN Ia, will serve the important purpose of either confirming the results derived from SN Ia or offering insights for improving the cosmological use of SN Ia.  Having an independent set of distances will enable a deeper study of SN Ia systematic errors and evolution with redshift.  A larger set of SN II light curves is also useful for efforts to better photometrically identify SN types.  By having a more complete picture of the range and frequency of SN II light curve properties, positive identification of SN II will be more likely, which will be useful both in building up a sample of SN II and in excluding them with confidence from samples of other types, particularly SN Ia.  This will become increasingly important as large surveys provide far more SN light curves than spectroscopic identification resources can handle.

\acknowledgments

We thank the FLWO staff for their dedicated work in maintaining the 1.2m and 1.5m telescopes and instruments, as well as the 1.3m PAIRITEL when it was in full operation.  We likewise thank the MMT staff.  We are grateful to Dan Scolnic for discussions on comparing the CfA SN II star sequences with the Pan-STARRS1 stars, and Carlos Contreras and Joseph Anderson for help in comparing CfA SN II star sequences with the CSP.  We also thank Kaisey Mandel and Arturo Avelino for helpful discussions.  This work has been supported, in part, by NSF grants AST-0606772, AST-0907903, AST-1211196, AST-1516854 and NASA grant NNX15AJ55G to Harvard University.  A.S.F. acknowledges support from NSF Awards SES 1056580 and PHY 1541160.  M. Modjaz is supported in part by the NSF CAREER award AST-1352405 and by the NSF award AST-1413260. 

\facilities{ FLWO:1.2m, FLWO: PAIRITEL, FLWO:1.5m, MMT}

\clearpage


\clearpage
\begin{thebibliography}{}

\bibitem[Anderson et al.(2014)]{anderson} Anderson, J. P., González-Gaitán, S., Hamuy, M., et al. 2014, ApJ, 786, 67
\bibitem[Arcavi et al.(2012)]{arcavi} Arcavi, I., Gal-Yam, A., Cenko, S. B., et al. 2012, ApJ, 756, L30
\bibitem[Bianco et al.(2014)]{Bianco14} Bianco, F. B., Modjaz, M., Hicken, M., et al. 2014, ApJS, 213, 19
\bibitem[Bloom et al.(2006)]{bloom06} Bloom, J. S., Starr, D. L., Skrutskie, M. F., et al. 2006, ASPC, 351, 751
\bibitem[Blondin et al.(2012)]{Blondin12} Blondin, S., Matheson, T., Kirshner, R. P., et al. 2012, AJ, 143, 126
\bibitem[Cutri et al.(2003)]{cutri03} Cutri, R. M., Skrutskie, M. F., van Dyk, S., et al. 2003, The IRSA 2MASS All-Sky Point Source Catalog,  NASA/IPAC Infrared Science Archive
\bibitem[de Jaeger et al.(2017)]{dejaeger} de Jaeger, T., González-Gaitán, S.,  Hamuy, M., et al. 2017, ApJ, 835, 166
\bibitem[D'Andrea et al.(2010)]{dandrea} D'Andrea, C. B., Sako, M., Dilday, B., et al. 2010, ApJ, 708, 661
\bibitem[Dessart et al.(2008)]{dessart08} Dessart, L., Blondin, S., Brown, P. J., et al. 2008, ApJ, 675, 644
\bibitem[Fabricant et al.(1998)]{Fabricant98} Fabricant, D., Cheimets, P., Caldwell, N. \& Geary, J. 1998, PASP, 110, 79
\bibitem[Faran et al.(2014)]{faran} Faran, T., Poznanski, D., Filippenko, A. V., et al. 2014, MNRAS, 442, 844
\bibitem[Fransson et al.(2014)]{fransson14} Fransson, C., Ergon, M., Challis, P., et al. 2014, ApJ, 797, 118
\bibitem[Friedman et al.(2012)]{friedman12} Friedman, A. S., PhD thesis, Harvard University
\bibitem[Friedman et al.(2015)]{Friedman15} Friedman, A. S., Wood-Vasey, W. M., G. H. Marion, et al. 2015, ApJS, 220, 9
\bibitem[Galbany et al.(2016)]{galbany} Galbany, L., González-Gaitán, S.,  Phillips, M. M., et al. 2016, AJ, 151, 33
\bibitem[Gall et al.(2017)]{gall17} Gall, E. E. E., Kotak, R., Leibundgut, B., et al. arXiv:1705.10806 [astro-ph.CO]
\bibitem[Gall et al.(2015)]{gall15} Gall, E. E. E., Polshaw, J., Kotak, R., et al. 2015, A\&A, 582, A3
\bibitem[Gal-Yam(2016)]{galyam} Gal-Yam, A. arXiv:1611.09353 [astro-ph.HE]
\bibitem[Gonz\'alez-Gait\'an et al.(2015)]{gonzalezgaitan} Gonz\'alez-Gait\'an, S., Tominaga, N., Molina, J., et al. 2015, MNRAS, 451, 2212
\bibitem[Guti\'errez(2016)]{gutierrez} Guti\'errez, C., (2016) \textit{Spectral Analysis of Type II Supernovae} (Doctoral Thesis).  Retrieved from http://repositorio.uchile.cl/handle/2250/141777
\bibitem[Hicken et al.(2009)]{Hicken09} Hicken, M., Wood-Vasey, W. M., Blondin, S., et al. 2009, ApJ, 700, 331
\bibitem[Hicken et al.(2012)]{Hicken12} Hicken, M., Challis, P., Kirshner, R. P., et al. 2012, ApJS, 200, 12
\bibitem[Jha et al.(2006)]{Jha06} Jha, S., Kirshner, R. P., Challis, P., et al. 2006, AJ, 131, 527
\bibitem[Kirshner \& Kwan(1974)]{kirshner74} Kirshner, R. P. \& Kwan, J. 1974, ApJ, 193, 27
\bibitem[Kirshner \& Kwan(1975)]{kirshner75} Kirshner, R. P. \& Kwan, J. 1975, ApJ, 197, 415
\bibitem[Liu et al.(2016)]{liu16} Liu, Y. Q., Modjaz, M., Bianco, F. B., \& Graur, O. 2016, ApJ, 827, 90
\bibitem[Maguire et al.(2010)]{maguire10} Maguire, K., Kotak, R., Smartt, S. J., et al. 2010 MNRAS, 403, L11 
\bibitem[Marion et al.(2014)]{marion14} Marion, G. H., Vinko, J., Kirshner, R. P., et al. 2014, ApJ, 781, 69
\bibitem[Matheson et al.(2008)]{Matheson08} Matheson, T., Kirshner, R. P., Challis, P., et al. 2008, AJ, 135, 1598
\bibitem[Matheson et al.(2005)]{Matheson05} Matheson, T., Blondin, S., Foley, R. J., et al. 2005, AJ, 129, 2352
\bibitem[Modjaz et al.(2014)]{Modjaz14} Modjaz, M., Blondin, S., Kirshner, R. P., et al. 2014, AJ, 147, 99
\bibitem[Modjaz et al.(2016)]{modjaz16} Modjaz, M., Liu, Y. Q., Bianco, F. B., \& Graur, O. 2016, ApJ, 832, 108
\bibitem[Morozova, Piro \& Valenti(2017)]{morozova17} Morozova, V., Piro, A. L., \& Valenti, S. 2017, ApJ, 838, 28
\bibitem[Pastorello et al.(2015)]{pastorello} Pastorello, A., Benetti, S., Brown, P. J., et al. 2015, MNRAS, 449, 1921 
\bibitem[Planck Collaboration(2016)]{planck16} Planck Collaboration:  Ade, P. A. R., Aghanim, M., Arnaud, M., et al. 2016, A\&A, 594, A13
\bibitem[Poznanski et al.(2009)]{poznanski09} Poznanski, D., Butler, N., Filippenko, A. V.,  et al. 2009, ApJ, 694, 1067
\bibitem[Richardson et al.(2002)]{richardson02} Richardson, D., Branch, D., Casebeer, D., et al. 2002, AJ, 123, 745
\bibitem[Riess et al.(1999)]{Riess99} Riess, A. G., Kirshner, R. P., Schmidt, B. B., et al. 1999, AJ, 117, 707
\bibitem[Riess et al.(2016)]{riess16} Riess, A. G., Macri, L. M., Hoffman, S. L., et al. 2016, ApJ, 826, 56
\bibitem[Rodr\'iguez et al.(2014)]{rodriguez14} Rodr\'iguez, O., Clocchiatti, A., \& Hamuy, M. 2014, AJ, 148, 107
\bibitem[Rodr\'iguez et al.(2016)]{rodriguez16} Rodr\'iguez, O., Pignata, G. \& Hamuy, M., 2016, http://sn2016.cl/documents/posters/poster\_rodriguez.pdf
\bibitem[Rubin et al.(2015)]{rubin} Rubin, A., Gal-Yam, A., De Cia, A., et al. arXiv:1512.00733 [astro-ph.HE]
\bibitem[Sanders et al.(2015)]{sanders} Sanders, N. E., Soderberg, A. M., Gezari, S., et al. 2015, ApJ, 799, 208
\bibitem[Schechter et al.(1993)]{schechter93} Schechter, P. L., Mateo, M., \& Saha, A. 1993, PASP, 105, 1342
\bibitem[Schmidt, Kirshner \& Eastman(1992)]{schmidt92} Schmidt, .B, P., Kirshner, R. P., \& Eastman, R., G. 1992, ApJ, 395, 366
\bibitem[Schmidt et al.(1994)]{schmidt94} Schmidt, B., P., Kirshner, R. P.,  Eastman, R., G., et al. 1994, ApJ, 432, 42
\bibitem[Scolnic et al.(2015)]{Scolnic15} Scolnic, D., Casertano, S., Riess, A., et al. 2015, ApJ, 815, 117 
\bibitem[Skrutskie et al.(2006)]{skrutskie06} Skrutskie, M. F., Cutri, R. M., Stiening, R., et al. 2006, AJ, 131, 1163
\bibitem[Smartt et al.(2009)]{smartt} Smartt, S. J., Eldridge, J. J., Crockett, R. M., \& Maund, J. R. 2009, MNRAS,
395, 1409
\bibitem[Smith et al.(2002)]{Smith} Smith, J. A., Tucker, D. L., Kent, S., et al. 2002, AJ, 123, 2121
\bibitem[Stritzinger et al.(2011)]{stritzinger} Stritzinger, M. D., Phillips, M. M., Boldt, L. N., et al. 2011, AJ, 142, 156 
\bibitem[Taddia et al.(2013)]{taddia} Taddia, F., Stritzinger, M. D., Sollerman, J., et al. 2013, A\&A, 555, A10
\bibitem[Tonry et al.(2012)]{Tonry12} Tonry, J. L., Stubbs, C. W., Lykke, K. R., et al. 2012, ApJ, 750, 99
\bibitem[Van Dyk et al.(2012)]{vandyk} Van Dyk, S. D., Cenko, S. B., Poznanski, D., et al. 2012, ApJ, 756, 131
\bibitem[Wood-Vasey et al.(2008)]{Woodvasey08} Wood-Vasey, W. M., Friedman, A. S., Bloom, J. S., et al. 2008, ApJ, 689, 377
\bibitem[Valenti et al.(2016)]{valenti16} Valenti, S., Howell, D. A., Stritzinger, M. D., et al. 2016, MNRAS, 459, 3939

\end{thebibliography}
\end{document}